\documentclass[prc,onecolumn,tightenlines,12pt]{revtex4}
\usepackage{graphicx}
\usepackage{dcolumn}
\usepackage{amsmath}
\newcommand{\be}{\begin{equation}}
\newcommand{\ee}{\end{equation}}
\newcommand{\bea}{\begin{eqnarray}}
\newcommand{\eea}{\end{eqnarray}}

\begin{document}

\author{J. X. de Carvalho$^{1,2}$, 
M. S. Hussein$^{\dag 1,2}$, 
M. P. Pato${^2}$ and A. J. Sargeant$^{2}$}
\affiliation{$^{1}$Max-Planck-Institut f\"ur Physik komplexer Systeme\\
N\"othnitzer Stra$\beta$e 38, D-01187 Dresden, Germany \\
$^{2}$Instituto de F\'{i}sica, Universidade de S\~{a}o Paulo\\
C.P. 66318, 05315-970 S\~{a}o Paulo, S.P., Brazil}
\title{Symmetry Breaking Study with  Deformed Ensembles
\thanks{Supported in part by the CNPq and FAPESP (Brazil).\\
        $^{\dag}$Martin Gutzwiller  Fellow, 2007/2008.}}

\begin{abstract}

A random matrix model to describe the coupling of $m$-fold symmetry 
is constructed. The particular threefold case is used to analyze 
data on eigenfrequencies of elastomechanical vibration of an 
anisotropic quartz block. It is suggested that such 
experimental/theoretical study may supply powerful means to
discern intrinsic symmetries of physical systems.
\end{abstract}

\maketitle

The standard ensembles of Random Matrix Theory
(RMT) \cite{Meht} have had wide application
in the description of the statistical properties of eigenvalues and
eigenfunctions of complex many-body systems. Other ensembles have
also been introduced \cite{Dyson}, in order to cover situations that depart
from universality classes of RMT. One such class of ensembles is the
so-called Deformed Gaussian Orthogonal Ensemble (DGOE)
\cite{Pato1,Pato2,Pato3, Carneiro:1991} that proved to be particularly useful when one
wants to study the breaking of a discrete symmetry in a
many-body system such as the atomic nucleus.

In fact, the use of spectral statistics as a probe of symmetries in physical
systems has been a subject of intensive experimental and theoretical
investigation following the pioneering work of Bohigas, Giannoni and Schmit 
\cite{Bohigas}
which showed that the quantal behaviour of classically chaotic systems exhibits the
predictions supplied by the RMT. Examples of symmetry breaking in physical
systems that have been studied include nuclei \cite{Mitch0, Mitch}, atoms 
\cite{Simons, Welch}
and mesoscopic devices such as quantum dots \cite{Alhassid}.

In the case of nuclei, the Mitchell group at the Triangle Universities Nuclear
Laboratory \cite{Mitch0, Mitch}, studied the effect of isospin symmetry breaking, 
in  odd-odd
nuclei such as $^{26}Al$. They detected the breakdown of this important symmetry by
the applications of two statistics: the short-range, nearest neighbor level
spacing distribution (NND) and the long range Dyson's $\Delta$-statistics 
\cite{Mitch0, Mitch}. These results were well described by a DGOE in which a pair of 
diagonal blocks is coupled. The strength of the coupling needed to account for the 
symmetry breaking can be traced to the average matrix element of the Coulomb interaction
responsible for this discrete symmetry breaking \cite{Pato2, Guhr}.  
The justification for the use of block matrices to describe the statistics of
a superposition of $R$ spectra with different values of the conserved
quatum number can be traced to Refs. \cite{Meht, NovaR}.
In the case of non-interacting
spectra, \emph{i.e.} if the quantum number is exactly conserved, the answer is
a superposition of the $R$ spectra. Since the level repulsion is present in
each one of the $R$ spectra, their superposition does not show this feature. 
Thus, we can say that for each spectra of states of a given value 
of the quantum number, one attaches a random matrix (GOE). For $R$ spectra
each of which has a given value of the conserved quantum number, one
would have an $R\times R$ block diagonal matrix. Each block matrix will 
have a dimension dictated by the number state of that spectra. If the
quantum number is not conserved then the $R\times R$ block matrix acquires non-diagonal
matrices that measure the degree of the breaking of the associated symmetry.
This idea was employed by Guhr and Weidenm\"uller \cite{Guhr} and Hussein
and Pato \cite{Pato1} to discuss isopin violation in the nucleus $^{26}$\emph{Al}.
In reference \cite{Pato1}, the random block matrix model was called
the Deformed Gaussian Orthogonal Ensemble (DGOE).

In order to study transitions amongst universal classes of ensembles
such as order-chaos (Poisson$\rightarrow$GOE), symmetry violation 
transitions (2GOE$\rightarrow$1GOE), experiments on physical 
systems are more complicated due to the difficulty of tuning the interaction
(except, e.g. in highly excited atoms where the application of a
magnetic field allows the study of GOE-GUE transitions). To simulate
the microscopic physical systems, one relies on analog computers such
as microwave cavities, pioneered by A. Richter and collaborators
\cite{achim1} and acoustic resonators of Ellegaard and 
collaborators\cite{Elleg0,Elleg,Elleg2}. It is worth mentioning at this
point that the first to draw attention to the applicability of RMT to
accoustic waves in physical system was Weaver \cite{Weaver}.

In the experiment of Ellegaard \emph{et al.} \cite{Elleg} what 
was measured were eigenfrequencies of the elastomechanical vibrations 
of an anisotropic crystal block with a D3 point-group symmetry.
The rectangular crystal block employed by Ellegard was so prepared as
to have only a two-fold flip symmetry retained. Then, to all
effects, the quartz specimen resembles a system of two three-dimensional Sinai
billiards. The statistical treatment of the eigenfrequencies of such a
block would follow that of the superposition of two uncoupled GOE's.

Then, 
by removing octants of  progressively larger radius from a corner of 
the crystal block this remnant two-fold symmetry was gradually broken. 
The spectral statistics show a transition towards 
fully a chaotic system as the octant radius increases. 
What was then seen was that the measured
NND is compatible with a two block DGOE description but the 
$\Delta$-statistics was discrepant. This discrepancy was attributed to 
pseudo integrable behavior and this explanation was later implemented
with the result that the long-range behavior was fitted at the cost, 
however, of loosing the previous agreement shown by the NND\cite{Abul}. 

Here we reanalyse this experiment following the simpler idea of
extending the DGOE matrix model \cite{Pato3} to consider the 
coupling of three instead of two GOE's \cite{Carneiro:1991}. 
We show that 
within this extension both, the short- and the long-range statistics,
are reasonably fitted suggesting that the assumption 
of the reduction of the complex symmetries of anisotropic quartz block
may not be correct. Our findings have the potential of supplying  very
precise means of testing details of symmetry breaking in pysical systems.

To define the ensembles of random matrices we are going to work with, we
recall the construction based on the Maximum Entropy Principle \cite{Pato1},
that leads to a random Hamiltonian which can be cast into the form
\begin{equation}
H=H_{0}+\lambda H_{1},  \label{eq 1b}
\end{equation}
where the block diagonal $H_{0}$ is a matrix made of $m$ uncoupled 
GOE blocks and 
$\lambda $ ($0\leq \lambda \leq 1)$ is the parameter that controls the
coupling among the blocks represented by the $H_{1}$ off-diagonal
blocks. For $\lambda=1,$ the $H_{1}$ part completes the  $H_{0}$ part 
and $H=H^{GOE}.$

These two matrices $H_{0}$ and $H_{1}$ are better expressed introducing the 
following $m$ projection operators 
\begin{equation}
P_{i}=\sum\limits_{j\in I_{i}} \mid j><j\mid,  \label{eq 2a}
\end{equation}
where $I_{i}$ defines the domain of variation of the row and column 
indexes associated with $i$th diagonal block of size $M_i.$
Since we are specifically interested in the transition from 
a set of $m$ uncoupled GOE's to a single GOE,
we use the above projectors to 
generalize our previous model \cite{Pato1,Pato2} by writing
\begin{equation}
H_{0}=\sum\limits_{i=1}^{m}P_{i}H^{GOE}P_{i}  \label{eq 4}
\end{equation}
and
\begin{equation}
H_{1}=\sum\limits_{i=1}^{m}P_{i}H^{GOE}Q_{i}  \label{eq 5}
\end{equation}
where $Q_{i}=1-P_{i}.$ It is easily verified that $H=H^{GOE}$
for $\lambda =1.$

The joint probability distribution of matrix elements can be put 
in the form \cite{Pato1,Alberto}
\begin{equation}
P(H,\alpha,\beta)=Z_{N}^{-1} \exp\left(-\alpha tr H^2 -
\beta tr H_{1}^2\right)  \label{eq 12a} 
\end{equation} 
with the parameter $\lambda$ being given in terms of $\alpha$ 
and $\beta$ by

\begin{equation}
\lambda=(1+\frac{\beta}{\alpha})^{-1/2}.  \label{eq 12b}
\end{equation}

Statistical measures of the completely uncoupled $m$ blocks have been 
derived. They show that level repulsion disappears which can be 
understood since eigenvalues from different blocks behave 
independently. In fact, as $m$ increases the Poisson statistics are
gradually approached. In the interpolating situation of partial 
coupling, some approximate analytical results have been derived.
In Ref. \cite{Alberto}, for instance, it has been found that the 
density $\rho (E)$ for arbitrary $\lambda$ and $m$ is given by
\begin{equation}
\rho(E)=\sum\limits_{i=1}^{m}\frac{M_i}{N}\rho_{i}(E)  \label{eq 5d}
\end{equation}
where 
\begin{equation}
\rho_{i}(E)=\left\{ 
\begin{array}{rl} 
\frac{2}{\pi a_{i}^{2}}\sqrt{{a_{i}^{2}-E^2}}, & \mid E\mid \leq a \\ 
0, &  \mid E \mid   > a
\end{array}
\right. 
\end{equation} \label{eq 5g}
is Wigner's semi-circle law with $a=\sqrt{N/\alpha}$ and
\begin{equation}
  a_{i}^{2}=a^{2}\left[\frac{M_{i}}{N}+\lambda^{2}\left(1-\frac{M_{i}}{N}\right)\right].
\end{equation}

Eq.~(\ref{eq 12a}) can be used to calculate exactly analytically 
the NND for $2\times 2$ and 
$3\times 3$ matrices \cite{Carneiro:1991}. For the $2\times 2$ case the
DGOE, Eq. (\ref{eq 12a}), gives 
\begin{equation}
  P_{2\times2}(s,\beta)=\alpha s\exp\left(-\frac{\alpha}{2}s^{2}\right)\sqrt{1+\frac{\beta}{\alpha}}I_{0}(\frac{\beta s^{2}}{4})
                    \exp\left(-\frac{\beta s^{2}}{4}\right),\label{rio}
\end{equation}
where $I_{0}$ is the modified Bessel function, whose asymptotic form is
\begin{equation}
     I_{0}(x) \rightarrow \frac{e^{x}}{\sqrt{2\pi x}}.
\end{equation}
Thus,there is no level repulsion for $\beta \rightarrow \infty$, $P_{2\times 2}(s,\infty)=\frac{2}{\pi}\alpha\exp\left(-\frac{\alpha}{2}s^{2}\right)$, which represents the 2x2 Poisson distribution where the usual exponential is replaced by a Gaussian. The prefactor is just 1 if 2$\alpha$ is taken to be $\pi$.
In the opposite limit, $\beta \rightarrow 0$, 
$I_{0}(x)\approx 1 - x^{2}/4$ and one obtains the Wigner distribution,
\begin{equation}
  P_{2\times 2}(s, \beta \rightarrow 0 )\approx \frac{\pi}{2}exp\left(-\frac{\pi}{4}s^{2}\right)s \label{mocambo}
\end{equation}
Note that the parameter $\lambda$ of eq (\ref{eq 12b}) is 0 if $\beta$ is $\infty$ and 1 if $\beta$ is 0.

For higher dimensions Eq.~(\ref{eq 12a}) can only be used for 
numerical simulations. This is what we are now reporting, using 2 and 3 bolck matrices of sizes 105 x 105 and 70 x 70 each,  respectively. The size of the whole matrix is 210 x 210. Further, we take an ensemble of 1000 elements and fix $\alpha$ to be 1.
We apply our model to analyse the eigenfrequency 
data of the elastomechanical vibrations of an anisotropic quartz 
block used in \cite{Elleg}. 

In this reference in order to break
the flip symmetry of the crystal block gradually they removed
an octant of a sphere of varying size at one of the corners.
The rectangular quartz block has the dimensions 
$14\times 25\times 40\, mm^{3}$. The radii of the spheres containing
the octants are $r=0.0, 0.5, 0.8, 1.1, 1.4$ and $1.7\, mm$ representing
figures $(a)-(f)$. Figs. $1x$ and $2x$ of Ref. \cite{Elleg} 
correspond to an octant of a huge sphere of radius $r=10.0\, mm$, whose center is inside the crystal and close to one of the corners.
They found  1424, 1414, 1424, 1414, 1424 and 1419 frequency eigenmodes in cases $(a)-(f)$,
respectively. These eigenfrequecies were measured in the frequency range between 600 and 900 kHz. Thus the average spacing between the modes is about 214Hz.
The histograms and circles
in the two figures of Ref. \cite{Elleg} represent the short-range
nearest-neighbor distributions $P(s)$ (Fig. 1) and the long range
$\Delta_{3}(L)$ statistics (Fig. 2).
In our DGOE simulation the unfolding of the calculated spectra is performed with the DGOE density given by Eq. (7) above.

                In figures 1 and 2, we show the 
results of our simlulations as compared to the data
of Ellegaard \emph{et al.} \cite{Elleg} for the spacing distribution and in figures 3 and 4  the long range 
correlation exemplified by the 
spectral rigidity $\Delta_{3}(L)$. We simulate the gradual breaking of the 2- or 3-fold symmetry by changing the value 
of the parameter $\lambda$ above. We see clearly that in so far as the $\Delta_{3}(L)$ is concerned a 3-GOE 
description works much better than a 2-GOE one. It is clear, however that both descriptions fall below the data, 
specially at large $L$. 
We shall analyse this discrepancy in the following using the missing level effect\cite{ML}.

It is often the case that there are some missing levels in
the statistical sample analysed. Such a situation was
addressed recently by Bohigas and Pato \cite{ML} who have
shown that if $g$ fraction of the levels or eigenfrequencies
is missing, the $\Delta_{3}(L)$ becomes
\begin{equation}
  \Delta_{3}^{g}(L) = g\frac{L}{15}+(1-g)^{2}\Delta_{3}
                                \left(\frac{L}{1-g}\right).\label{dragao}
\end{equation}
The presence of the linear term, even if small, could explain
the large $L$ behavior of the \emph{measured} $\Delta_{3}(L)$.
We call this effect the Missing Level (ML) effect. Another
possible deviation of $\Delta_{3}$  could arise
from the presence of pseudo-integrable effect (PI) \cite{Abul,bis2}.
This also modifies $\Delta_{3}$ by adding a Poisson term 
just like Eq. (\ref{dragao}). In the following we show that there is no need
for the PI effects to explain the large-L data on the $\Delta_{3}$ if the ML effect 
is taken into account.

We take a  study case Figs. 3b and 4b  which correspond to $r= 0.5\, mm$
and  where $1414$ frequency eigenvalues were found. We consider
this a potential ML  case and take for $\Delta_{3}$, the
expression given in Eq. (\ref{dragao}) and apply to our simulations. 
We find perfect fit to the \emph{data}, if $g$
is taken to be $0.1$, namely only $90\%$ of the eigenfrequencies
were in fact taken into account in the statistical analysis.
There is, threfore, room to account much better for all cases
(Fig. $2a$, $2c$, $\ldots$ ) by appropiately choosing the 
correponding value of $g$. We have also
verified that if a 2GOE description is used, namely, $m=2$ , then an
account of the large-$L$ behaviour of $\Delta_{3}$ can also be obtained if a
much larger number of levels were missing in the sample. In our
particular case of Fig. 2b, we obtained $g = 0.3$. This is 3 times larger
than the ML needed in the 3GOE description. We consider the large value
of $g$ needed in the 2GOE description, much too large
to conform to the reported data in Ref \cite{Elleg}. Figure 5 summarizes
our the above. 

It is therefore clear that the 3GOE description of the
spectral rigidity of the eifenfrequency spectra of \cite{Elleg} for the 
crystal block does work very well if a small fraction of the levels is 
taken to be missing, without resort to pseudointegrable trajectories or
levels that do not feel the symmetry breaking \cite{Abul}. On the other hand,
the 2GOE description, which does as good as the 3GOE one in fitting the measured 
$P(s)$, fails dramatically in accounting for the spectral rigidity, even if as
much as 30 per cent of the levels are taken as missing.

In conclusion, a random matrix model to describe the coupling 
of $m$-fold symmetry 
is constructed. The particular threefold case is used to analyse 
data on eigenfrequencies of elastomechanical vibration of a 
anisotropic quartz block. By properly taking into account the ML 
effect we have shown that the quartz block could very well be
described by 3 uncoupled GOE's , which are gradually coupled by the 
breaking of the three-fold symmetry
(through the gradual removal of octants of increasing sizes), till a
1GOE situation is attained. This, therefore, indicates that the
unperturbed quartz block may possess another symmetry, besides the flip one.
A preliminary version of the formal aspect of this work 
has previously appeared in \cite{last}.

\newpage

%%%%%%%%%%%%%%%%%%%%%%%%%%%%
\begin{figure}[h]
\includegraphics[width=\textwidth]{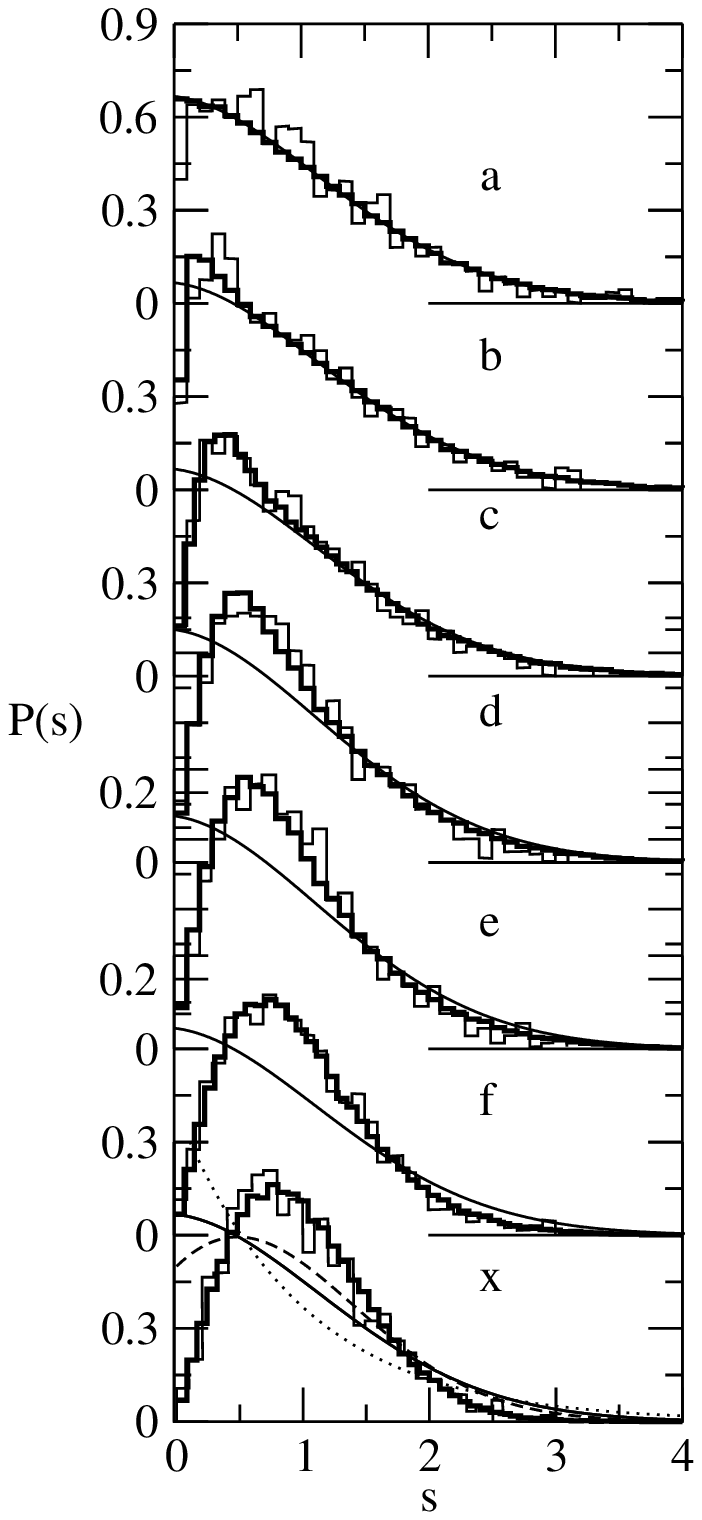}% Here is how to import EPS art
\caption{
Nearest Neighbour Distributions. Histograms show data (a)-(x) from
Ref. \cite{Elleg}. Thick histograms show the three coupled GOE  fits to the data
carried out using the DGOE numerical simulations using Eq. (\ref{eq 12a}). Also
shown as the full thin line  the three uncoulped GOE $P(s)$. In
graph (x) the dotted line is the Poisson distribution, the dashed line is the two uncoupled
GOE $P(s)$. The very thin line
is Wigner distribution which is hidden behind histograms.
The values of $\lambda$ that adjust the data are $0.0032, 0.0071, 0.0158, 0.0250, 0.0333, 0.9950, 1.000$
for cases (a)-(x). See text for details.
} \label{eterno1}
\end{figure}
%%%%%%%%%%%%%%%%%%%%%%%%%%
%%%%%%%%%%%%%%%%%%%%%%%%%%%%
\begin{figure}[h]
\includegraphics[width=\textwidth]{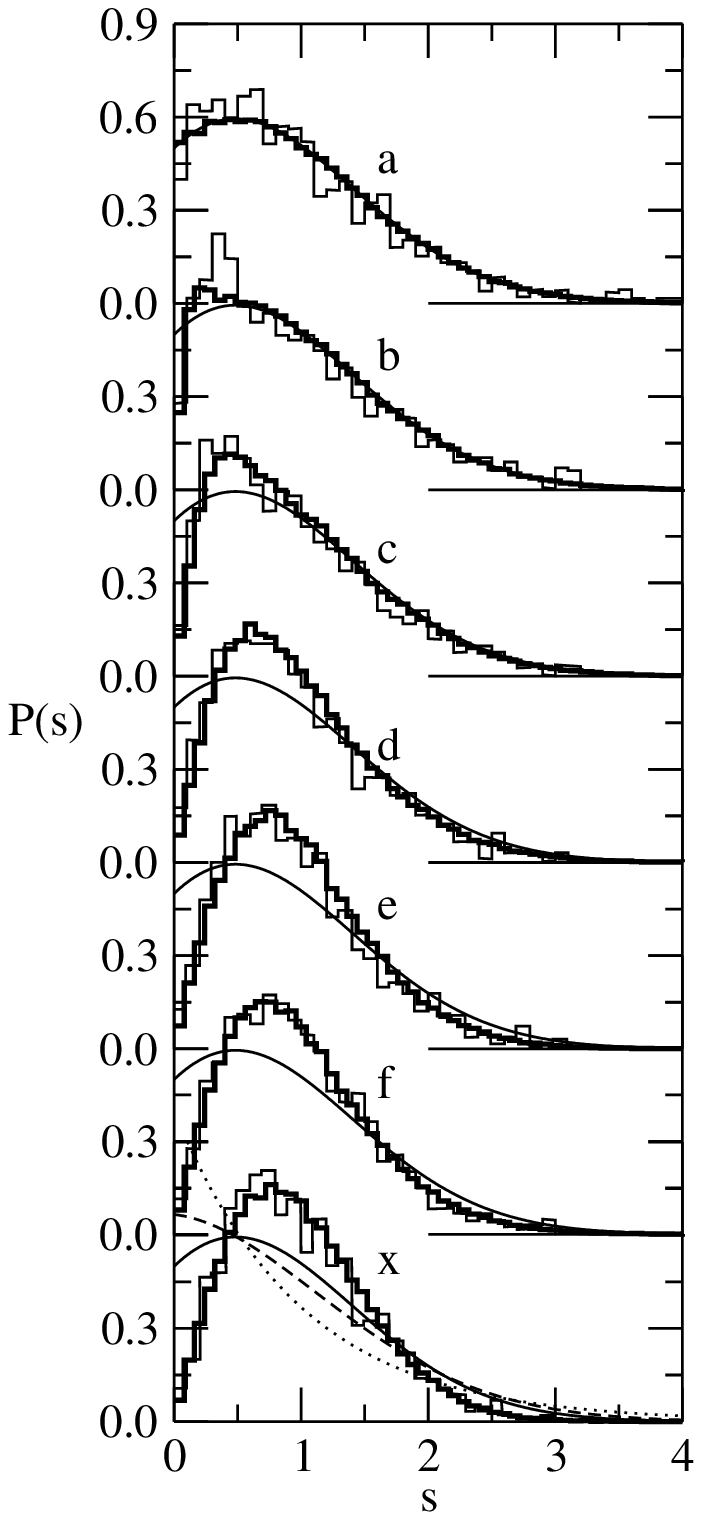}% Here is how to import EPS art
\caption{
Nearest Neighbour Distributions. Histograms show data (a)-(x) from
Ref. \cite{Elleg}. Thick histograms show the two coupled GOE  fits to the data
carried out using the DGOE numerical simulations using Eq. (\ref{eq 12a}). Also
shown as the full thin line  the two uncoulped GOE $P(s)$. In
graph (x) the dotted line is the Poisson distribution, the dashed line is three  uncoupled
GOE $P(s)$. The very thin line
is Wigner distribution which is hidden behind histograms.
The values of $\lambda$ that adjust the data are $0.000, 0.0258, 0.0200, 0.0400, 0.0705, 0.0600, 1.000$
for cases (a)-(x). See text for details.
} \label{eterno2}
\end{figure}
%%%%%%%%%%%%%%%%%%%%%%%%%%
\begin{figure}
\includegraphics[width=\textwidth]{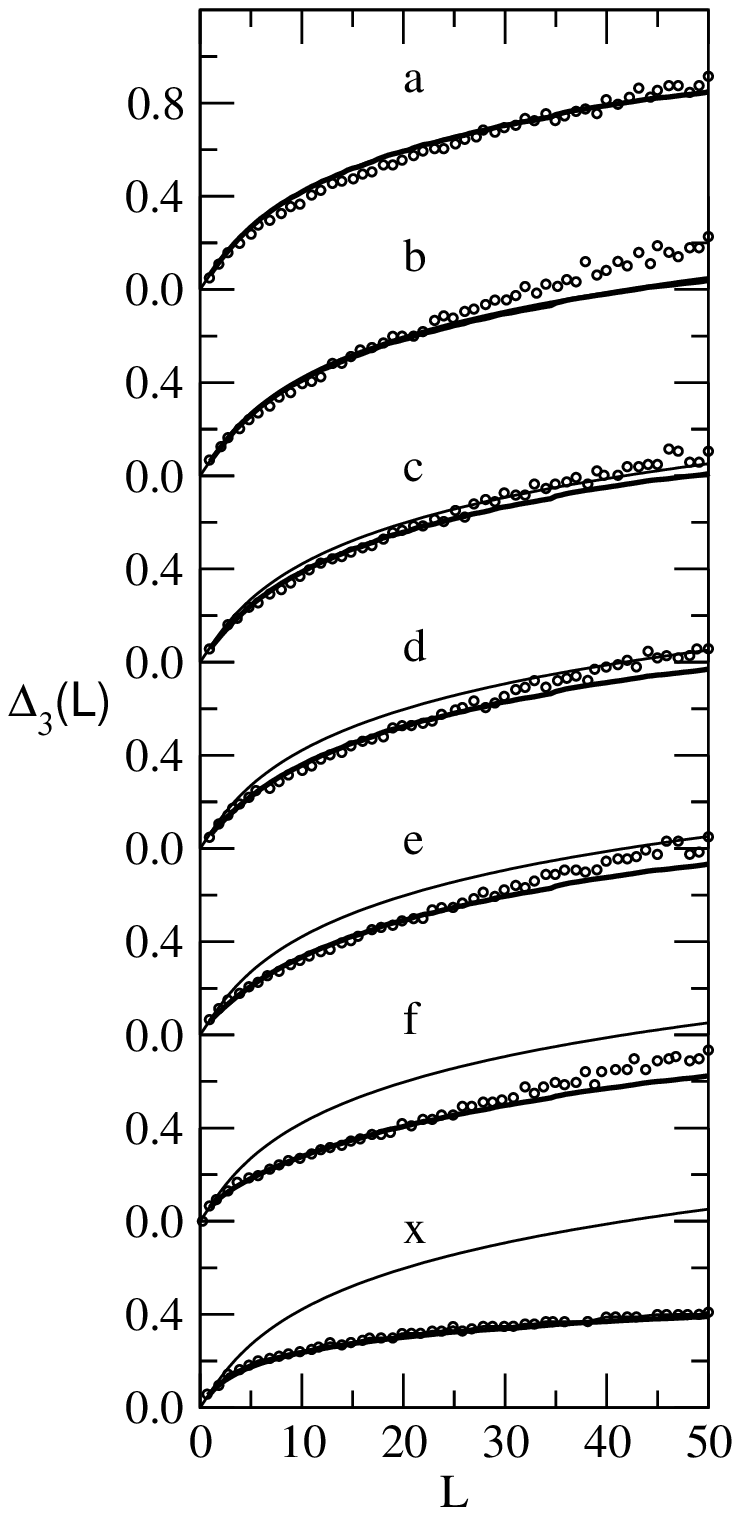}% Here is how to import EPS art
\caption{
Spectral Rigidities. The thick lines are the DGOE simulation for the three 
coupled GOE's.The same values of $\lambda$ as in Fig. 1 were used. 
The thin lines correspond the three uncoupled GOE's case.
The data points are from Ref. \cite{Elleg}. See text for details.
} \label{presente1}
\end{figure}
%%%%%%%%%%%%%%%%%%%%%%%%%%%%
\begin{figure}
\includegraphics[width=\textwidth]{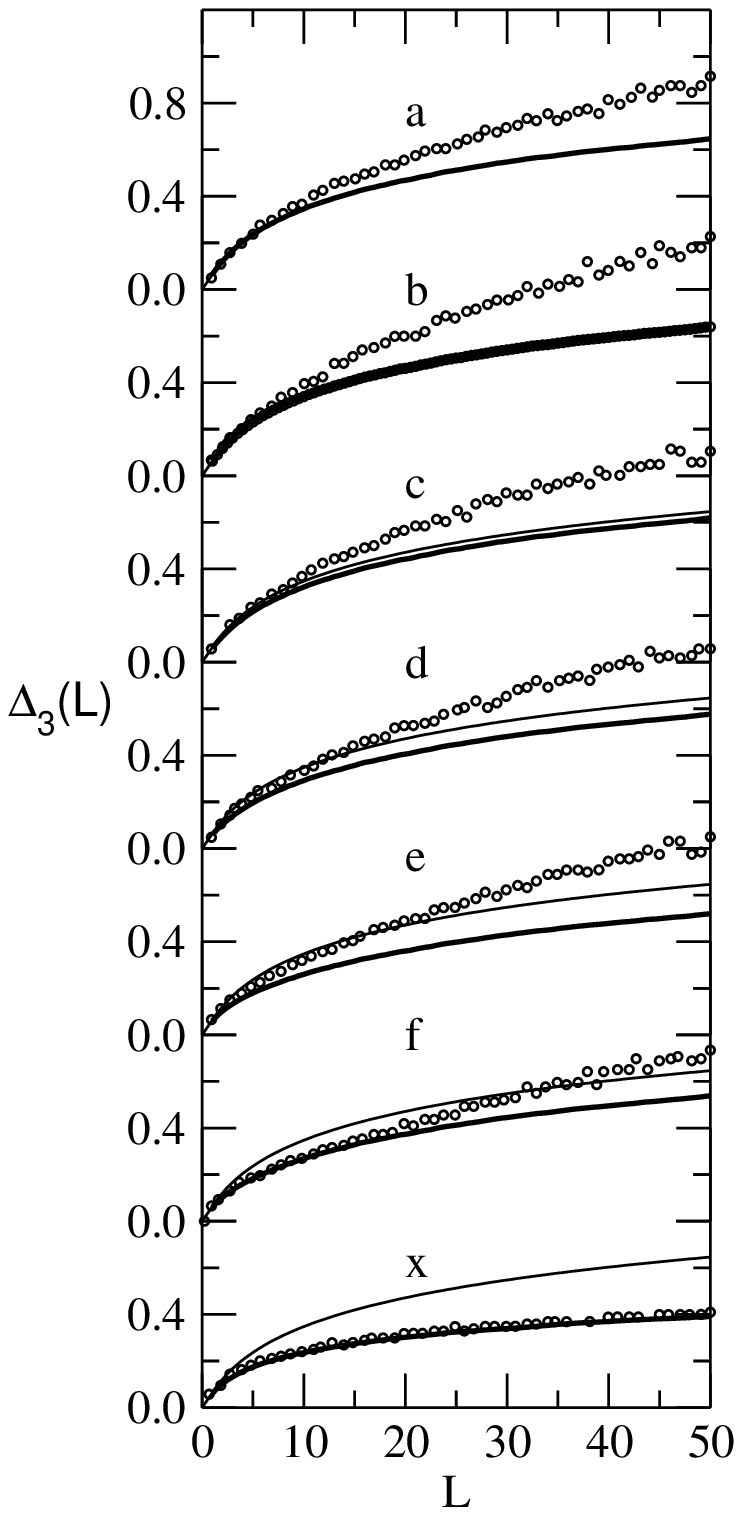}% Here is how to import EPS art
\caption{
Spectral Rigidities. The thick lines are the DGOE simulation for the two 
coupled GOE's. The same values of $\lambda$ as in Fig. 2 were used.
The thin lines correspond the two uncoupled GOE's case.
The data points are from Ref. \cite{Elleg}. See text for details.
} \label{presente2}
\end{figure}
%%%%%%%%%%%%%%%%%%%%%%%%%%%%
%%%%%%%%%%%%%%%%%%%%%%%%%%%%
\begin{figure}
\includegraphics[width=\textwidth]{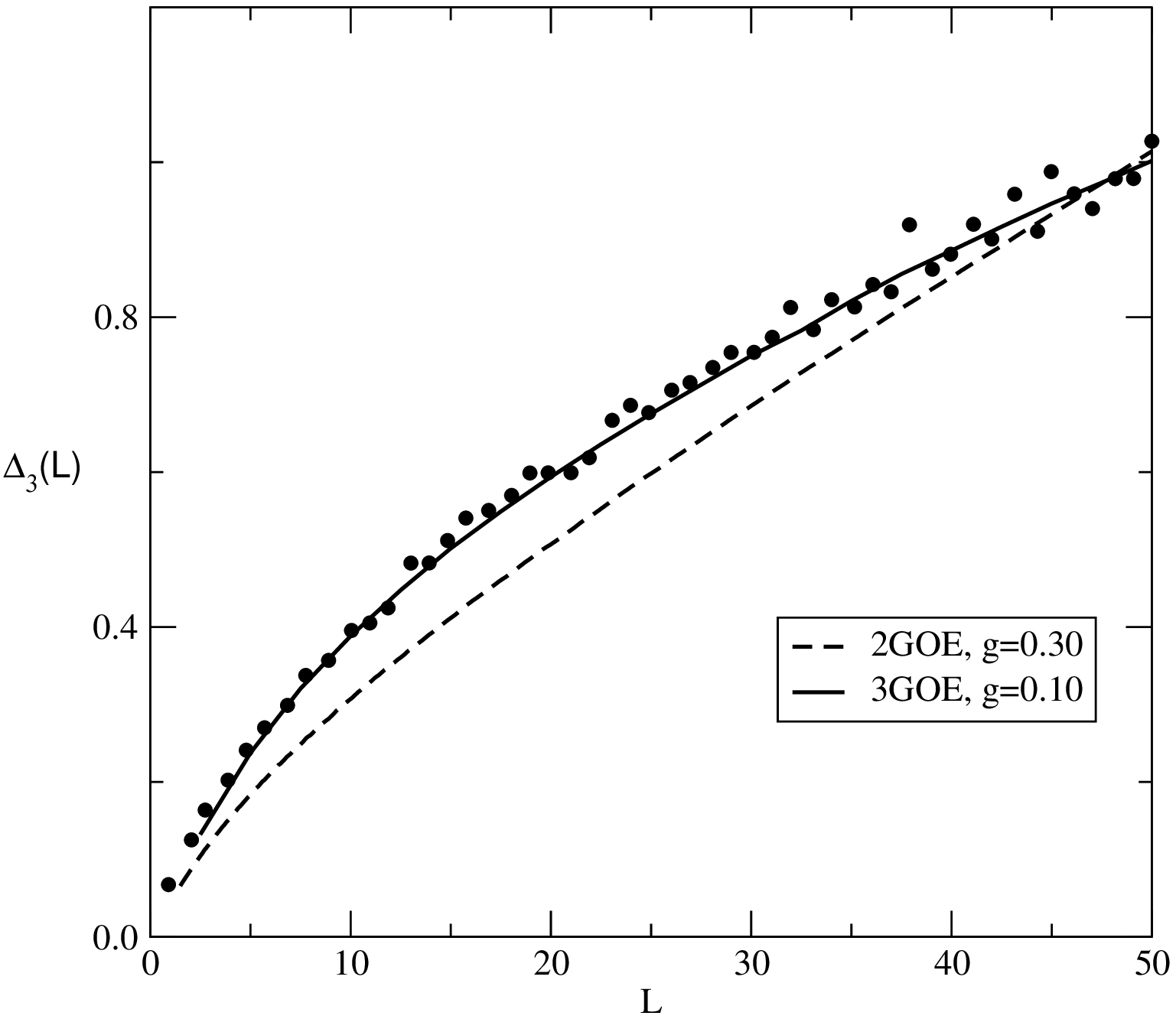}% Here is how to import EPS art
\caption{The ML effect. The data points correspond to case (b) of Ref. \cite{Elleg},
$r=0.5\,mm$. The full line corresponds to our three coupled GOE's fit with $\lambda=0.0071$,
figure 3b 
and $g=0.10$. The dashed line corresponds to our two coupled GOE's fit with $\lambda=0.0258$,
figure 4b
and $g=0.30$. See text for details.
} \label{presente3}
\end{figure}
%%%%%%%%%%%%%%%%%%%%%%%%%%%%


\newpage

\begin{thebibliography}{101}



\bibitem{Meht}  M.L. Mehta, {\it Random Matrices} 2nd Edition (Academic
Press, Boston, 1991); T.A. Brody et al., Rev. Mod. Phys.{\bf \ 53}, 385 (1981);
T. Guhr, A. M\"{u}ller-Groeling
and A. Weidenm\"{u}ller, Phys. Rep. {\bf 299, }189 (1998).

\bibitem{Dyson}  F.J. Dyson, J. Math. Phys. {\bf 3}, 1191 (1962).

\bibitem{Pato1}  M. S. Hussein, and M. P. Pato, Phys. Rev. Lett. {\bf 70, }

1089{\bf \ }(1993).

\bibitem{Pato2}  M. S. Hussein, and M. P. Pato, Phys. Rev. {\bf C 47,} 2401

(1993).

\bibitem{Pato3}  M. S. Hussein, and M. P. Pato, Phys. Rev. Lett. {\bf 80, }
1003{\bf \ }(1998).


\bibitem{Carneiro:1991}
C.~E. Carneiro, M.~S. Hussein and M.~P. Pato, in 
H.~A. Cerdeira, R.~Ramaswamy, M.~C. Gutzwiller and
  G.~Casati (eds.) \emph{Quantum Chaos}, p. 190 (World
  Scientific, Singapore) (1991).

\bibitem{Bohigas}O. Bohigas, M. J. Giannoni and C. Schmit, Phys. Rev. Lett.
\textbf{52}, 1 (1984). See also O. Bohigas and M. J. Giannoni, in 
Mathematical and Computational Methods in Nuclear Physics, edited by
J. S. DeHesa, J. M. Gomez and A. Polls, Lecture Notes in Physics Vol.
209 (Springer-Verlag, New York, 1984).


\bibitem{Mitch0}  G. E. Mitchell, E. G. Bilpuch, P. M. Endt, and F. J. Shriner,
Phys. Rev. Lett. {\bf 61}, 1473 (1988).

\bibitem{Mitch}  A.A. Adams, 
G.E. Mitchell, and J.F. Shriner, Jr., Phys. Lett. {\bf B 422,} 13 (1998).


\bibitem{Simons} B. D. Simons, A. Hashimoto, M. Courtney, D. Kleppner and B. L. Altshuler,
Phys. Rev. Lett. 71, 2899 (1993).

\bibitem{Welch} G. R. Welch, M. M. Kash, C-h Iu, L. Hsu and D. Kleppner, Phys. Rev. Lett.
62, 893 ( 1989).

\bibitem{Alhassid} See, e.g. Y. Alhassid, Rev. Mod. Phys. 72, 895 (2000) and references
therein.

\bibitem{Guhr}  T. Guhr and, H.A. Weidenm\"{u}ller. Ann. Phys. (NY), 
{\bf 199}, 412 (1990).


\bibitem{NovaR} N. Resenzweig and C. E. Porter, Phys. Rev. \textbf{120},
                1698 (1960).

\bibitem{achim1} H.-D. Graf, H. L. Harney, H. Lengeler,
C. H. Lewnkopf, C. Rangacharyulu, A. Richter, P. Schardt and H. A. 
Weidenmuller, Phys. Rev. Lett. 69, 1296(1992);
%
H. Alt, H. -D. Graf, H. L. Harney, R. Hofferbert, 
H. Lengeler, A. Richter, P. Schardt and H. A. Weidenmuller, 
Phys. Rev. Lett. 74, 62 (1995);
%
H. Alt, C. Dembowski, H. -D. Graf, R. Hofferbert, 
H. Rehfeld, A. Richter, R. Schuhmann and Weiland, Phys. Rev. 
Lett. 79, 1029 (1997);
%
H. Alt, C. I. Brabosa, H. -D. Graf, T. Guhr, H. L. 
Harney, R. Hofferbert, H. Rehfeld and A. Richter, Phys. Rev. Lett. 81,
4847 (1998);
%
C. Dembowski, H. -D. Graf, A. Heine, R. Hofferbert, H. Rehfeld and A.
Richter, Phys. Rev. Lett. 84, 867 (2000);
%
C. Dembowski, H. -D. Graf. H. L. Harney, A. Heine, W. D. Heiss, H.
Rehfeld and A. Richter, Phys. Rev. Lett. 86, 787 (2001);
%
C. Dembowski, H. -D. Graf, A. Heine, T. Hesse, H. Rehfeld and A. Richter,
Phys. Rev. Lett. 86, 3284 (2001);
%
C. Dembowski, B. Dietz, H. -D. Graf, A. Heine, T. Papenbrock, A. Richter
and C. Richter, Phys. Rev. Lett. 89, 064101-1 (2002);
%
C. Dembowski, B. Dietz, A. Heine, F. Leyvraz, M. Miski-Oglu, A. Richter
and T. H. Seligman, Phys. Rev. Lett. 90, 014102-1 (2003);
%
C. Dembowski, B. Dietz, H. -D. Graf, H. L. Harney, A. Heine, W. D. Heiss
and A. Richter, Phys. Rev. Lett. 90, 034101-1 (2003);
%
C. Dembowski, B. Dietz, T. Friedrich, H. -D. Graf, A. Heine, C.
Mejia-Monasterio, M. Miski-Oglu, A. Richter and T. H. Seligman, Phys. Rev.
Lett. 93, 134102-1 (2004);
%
B. Dietz, T. Guhr, H. L. Harney and A. Richter, Phys. Rev. Lett. 96,
254101 (2006);
%
E. Bogomolny, B. Dietz, T. Friedrich, M. Miski-Oglu, A. Richter, F.
Schafer and C. Schmit, Phys. Rev. Lett. 97, 254102 (2006);
%
 B. Dietz, T. Friedrich, H. L. Harney, M. Miski-Oglu, A. Richter, F.
Schafer and H. A. Weidenmuller, Phys. Rev. Lett. 98, 074103 (2007).


\bibitem{Elleg0} C. Ellegaard, T. Guhr, K. Lindemann, H. Q. Lorensen, J. Nygard and M.
Oxborrow, Phys. Rev. Lett. 75, 1546 (1995).

\bibitem{Elleg} C. Ellegaard, T. Guhr, K. Lindemann, J. Nygard and M. Oxborrow, Phys.
Rev. Lett. 77, 4918 (1996).

\bibitem{Elleg2} P. Bertelsen, C. Ellegaard, T. Guhr, M. Oxborrow and K. Schaadt, Phys.
Rev.Lett. 83, 2171 (1999).

\bibitem{Weaver} R. L. Weaver, J. Acoustic. Soc. Am. \textbf{85}, 1005 (1989).


\bibitem{Alberto} A. C. Bertuola, J. X. de Carvalho, M. S. Hussein, M. P. Pato, and 
A. J. Sargeant, Phys. Rev. E {\bf 71}, 036117 (2005).


\bibitem{ML} O. Bohigas and M. P. Pato, Phys. Lett. B,
             \textbf{595}, 171 (2004).


\bibitem{bis2} D. Biswas and S. R. Jain, Phys. Rev. A
               \textbf{42}, 3170 (1990).

\bibitem{Abul} A. Abd El-Hady, A. Y. Abul-Magd, and M. H. Simbel,
J. Phys. A {\bf 35}, 2361 (2002).


\bibitem{last} M. S. Hussein, J. X. de Carvalho, M. P. Pato and
               A. J. Sargeant, Few-Body Systems, \textbf{38},
               209 (2006).

\end{thebibliography}
\end{document}